\documentclass[prc,twocolumn]{revtex4}
\usepackage{epsfig}
\begin{document}
\flushbottom

\title{Higher order QED calculation of ultrarelativistic heavy ion
production of $\mu^+ \mu^-$ pairs}
\author{A. J. Baltz}
\affiliation{Physics Department,
Brookhaven National Laboratory,
Upton, New York 11973}
\date{\today}

\begin{abstract}
A higher order QED calculation of the ultraperipheral heavy ion cross section
for $\mu^+ \mu^-$ pair production at RHIC and LHC is carried out.
The so-called ``Coulomb corrections'' lead to an even greater
percentage decrease of $\mu^+ \mu^-$ production from perturbation theory
than the corresponding decrease for $e^+ e^-$ pair production.  Unlike the
$e^+ e^-$ case, the finite charge distribution of the ions (form factor)
and the necessary subtraction of impact parameters with matter overlap are
significant effects in calculation an observable ultraperipheral
$\mu^+ \mu^-$ total cross section.
\\
{\bf PACS: { 25.75.-q, 34.90.+q}}
\end{abstract}
\maketitle
\section{Introduction}

In recent years sufficient progress has been made in evaluating higher order
QED corrections to $e^+ e^-$ pair production in ultraperipheral heavy
ion collisions to allow a meaningful comparison with data.
The comparison of calculated $e^+ e^-$ pair production
rates\cite{ajb08} with STAR data\cite{star04} provides the first evidence of
higher order QED effects at RHIC.
The analogous higher order corrections to $\mu^+ \mu^-$ pair
production are now of interest with the anticipated program of ultraperipheral
heavy ion collisions at the LHC\cite{upc}.  In particular it has been
suggested by Kharlov and Sadovsky (see \cite{upc}) that since muon pairs
are easy to detect (in the ALICE detector) and simple to calculate
(in perturbation theory) they can be used as a luminosity
monitor at the LHC.  And a recent paper by Hencken, Kuraev, and
Serbo has presented approximate analytical calculations indicating that
the higher order Coulomb corrections to $\mu^+ \mu^-$ pair production are
small\cite{hks}.  However, in this paper I will present less approximate
numerical calculations showing that higher order QED corrections to the
$\mu^+ \mu^-$ rates are not small (and thus would require inclusion e.g.
in calculations to obtain a meaningful luminosity
measurement).  I will further argue that an additional reduction in
$\mu^+ \mu^-$ rates, relatively insignificant in the analogous $e^+ e^-$
process, arises from unitarity corrections necessitated by the exclusion of
ion trajectories with nuclear overlap.

In section II I discuss the method utilized here to calculate
higher order QED lepton pair production with heavy ions.  Section III
considers the differences between rates for $e^+ e^-$ and $\mu^+ \mu^-$
pair production in the usual impact parameter independent calculation:
how they scale and the relatively larger effect of the heavy ion form factor
for the $\mu^+ \mu^-$ pairs.  Section IV introduces the impact parameter
dependent representation necessary for consideration of unitarity effects,
especially the exclusion of ion overlap trajectories.  A summary of the 
results for computed total $\mu^+ \mu^-$ pair production cross sections at RHIC
and LHC is presented in Section V.

\section{Higher order QED: method of calculation}

The method employed in this paper for higher order $\mu^+ \mu^-$ pair
production is an extension of previous work on higher
order $e^+ e^-$ pairs.  A broad review of issues involved in the
$e^+ e^-$ pair calculations has been presented in a recent review\cite{brep},
and a discussion of the factorization of the different electromagnetic
processes and the applicability of the semiclassical description is found
in Ref.\cite{bhatk}.
Here I review the history relevant to the present calculations.

The possibility of accurate higher order $e^+ e^-$ pair
calculation originated with the realization that in an appropriate
gauge\cite{brbw},
the electromagnetic potential of a relativistic heavy ion is to a very good
approximation a delta function in the direction of motion of the heavy ion
times the two dimensional solution of Maxwell's equations in the transverse
direction\cite{ajb1,ajb2}.  This led to
the closed form solution of the time-dependent Dirac equation
for lepton pair production\cite{sw1,bmc,sw2}.  However this original
solution needed to be corrected for the following reason.  One apparent
consequence of the original solution was
that rates for pair production in the exact solution agreed with the 
corresponding perturbation theory result.  It was subsequently pointed out
by Ivanov,
Schiller, and Serbo\cite{serb} that this heavy ion conclusion was contrary
to the well known fact that photoproduction of $e^+ e^-$ pairs on a
heavy target shows a negative (Coulomb) correction proportional to $Z^2$
that is well described by the Bethe-Maximon theory\cite{bem}.
These authors went on to compute large Coulomb corrections to the pair total
cross section by considering higher order Feynman
diagrams in a leading logarithm approximation.
Lee and Milstein\cite{lm1,lm2} came to essentially the same result
for the Coulomb corrections.  They pointed out that the original
Dirac equation solution involved an integral over the
transverse spatial coordinates that was not well regularized.
Lee and Milstein constructed an appropriate regularized transverse
integral in the low transverse momentum ($k$) approximation that could be
solved analytically
to obtain the Coulomb corrections.  They also noted that replacing
the original transverse potential $- 2 i Z\alpha \ln(\rho)$ with
$2 i Z\alpha K_0(\rho \omega / \gamma)$ gives a properly regularized
expression for the original transverse integral
\begin{equation}
F({\bf k}) = 2 \pi \int d \rho \rho J_0(k \rho)
\{\exp [2i Z\alpha K_0(\rho \omega / \gamma)] -1 \} ,
\end{equation}
that goes over into the correct lowest order expression
\begin{equation}
F_0({\bf k}) = {4 i \pi Z\alpha \over k^2 + \omega^2/\gamma^2} .
\end{equation}
in the perturbative limit.
The modified Bessel function $K_0(\rho \omega / \gamma) = - \ln(\rho)$ plus
constants for small $\rho$
and cuts off exponentially at $\rho \sim \gamma / \omega $, where $\gamma$ is
the relativistic boost of the ion producing the photon and $\omega$ is the
energy of the photon.  I previously carried out numerical
calculations utilizing these expressions and obtained results identical to
those of Lee and Milstein in their small $k$ limit\cite{ajb3}.
In my previous cross section calculations\cite{ajb05,ajb06} and in what
follows, Eq. (1) is utilized for the higher order calculations
and Eq. (2) for lowest order.

For impact parameter dependent cross sections the calculations presented
here make use of the methods of calculating $e^+ e^-$ pair
probabilities previously described\cite{ajb06}.
The impact parameter ($b$) dependent amplitude presents a particular numerical
challenge since it involves a rapidly oscillating phase
$\exp(i \bf{k \cdot b})$ in the integral over the transverse momentum $\bf{k}$
transfered from the ion to the lepton pair.  The usual method of evaluating
the perturbative impact parameter dependent probability is to first square
the amplitude and then integrate over the sum and difference of
$\bf{k}$ and $\bf{k^\prime}$.  Here I have integrated before squaring, and
I deal with the rapid oscillations with the piecewise analytical integration
method previously described\cite{ajb06}.
In that previous $b$-dependent calculation of the total cross section
for $e^+ e^-$ production, half of
the contribution comes from $b > 5000$ fm and contributions up to $b = 10^6$
fm are considered.  Due to the large values of $b$ contributing, that
calculation was somewhat crude.  However integration over $b$ reproduced the
known cross sections calculated with the $b$ independent method or calculated
from the very accurate analytical Racah formula\cite{rac} to about 3\%.
It can also be noted that the computed perturbative $b$ dependent
probabilities in that paper were in relatively good agreement with
calculations in the literature\cite{hbt1} available for $ b < 7000 $~fm.

\section{Scaling of $\mu^+ \mu^-$ with $e^+ e^-$ Cross sections}

Let us begin by reviewing the scaling of $\mu^+ \mu^-$ cross sections
from the corresponding $e^+ e^-$ cross sections.
For point charge heavy ions (no form factor) if length is expressed
in terms of $1 / m_l$ and energy in terms of $m_l$ then the total
lepton pair cross section
$\sigma(\mu^+ \mu^-)$ is identical to $\sigma(e^+ e^-)$

A form factor $g(k)$ may be defined that modifies the expressions
for $F(k)$ in Eqns. (1) and (2). If one assumes a simple form factor
\begin{equation}
g(k) = {1 \over 1 + k^2/\Lambda^2}
\end{equation}
where for Au or Pb 
\begin{equation}
\Lambda \simeq 80\ {\rm MeV} = 160\ m_e = .75\ m_{\mu},
\end{equation}
then Eq. (2) for the perturbative limit becomes
\begin{equation}
F_0^{f}({\bf k}) = {4 i \pi Z\alpha \over (k^2 + \omega^2/\gamma^2) 
(1 + k^2/\Lambda^2)} .
\end{equation}
k is cut off at the low end when $k^2 \ll (\omega/\gamma)^2$.  
In the perturbative case for $e^+ e^-$ pairs it has been shown that the effect
of the form factor seems to be present only where the impact parameter is of
the same size as the nuclear radius\cite{htb}.  However the situation is
different for $\mu$ pairs.  At the high end the form factor cuts off when
$k^2 \gg \Lambda^2$.  The form factor contributes if
$\Lambda^2$ is comparable to or less than $(\omega/\gamma)^2$,
the cutoff of $k$ without the form factor.  Assume that at this high end
cutoff without the form factor 
$k \simeq 100 \omega/\gamma \simeq \omega$ for RHIC.  Clearly
for $\mu$ pair production the sum of the $\omega$s for the two virtual photons
must be greater than twice the mass of the muon.  Thus for even the lowest
energy $\mu$ pairs (corresponding to large
impact parameters) at least one $\omega > m_{\mu}$
and the form factor is important.
On the other hand, the form factor is relatively insignificant for
the total $\sigma(e^+ e^-)$ and contributes only at electron energies
some two orders of magnitude above the electron mass,
comparable to the value of $\Lambda$.
Without a form factor
\begin{equation}
{\sigma(\mu^+ \mu^-) \over \sigma(e^+ e^-)} =
\bigl({m_e \over m_{\mu}}\bigr)^2 = 2.34 \times 10^{-5}.
\end{equation}
But with a form factor the perturbation theory result calculations give
\begin{equation}
{\sigma(\mu^+ \mu^-) \over \sigma(e^+ e^-)} =
0.61 \times 10^{-5} = .26 \times \bigl({m_e \over m_{\mu}}\bigr)^2
\end{equation}
for RHIC, and
\begin{equation}
{\sigma(\mu^+ \mu^-) \over \sigma(e^+ e^-)} =
1.16 \times 10^{-5} = .50 \times \bigl({m_e \over m_{\mu}}\bigr)^2
\end{equation}
for LHC.  

To include a form factor in the eikonalized expression with
Coulomb corrections Eq. (1) then the most obvious prescription is to apply
the form factor to the transverse potential:
\begin{equation}
F^f({\bf k}) = 2 \pi \int d \rho \rho J_0(k \rho)
\{\exp [2i Z\alpha g(k) K_0(\rho \omega / \gamma)] -1 \} .
\end{equation}
This expression obviously goes into the correct perturbative limit
Eq. (5).  A simpler expression is to take the form factor only to
first order but the Coulomb corrections to higher order
\begin{equation}
F^{f0}({\bf k}) = 2 \pi g(k) \int d \rho \rho J_0(k \rho)
\{\exp [2i Z\alpha K_0(\rho \omega / \gamma)] -1 \} .
\end{equation}
Again, this expression obviously goes to the correct perturbative limit
Eq. (5).  This is the expression that will be used in this paper.  A discussion
of the validity of this approximation is given in Appendix A.

For simplicity in calculation and simplicity in comparing with Ref.\cite{hks},
the form factor $g(k)$ Eq. (3) has also neglected any dependence on
longitudinal momentum.  Including a longitudinal momentum dependence would
make a small reduction in cross section values, about 5\% for RHIC and 1\%
for LHC, as discussed in Appendix B.

I have previously calculated\cite{ajb05} that there is a 17\% reduction
at RHIC and a 11\% reduction at LHC
in the exact total $\sigma(e^+ e^-)$ from the perturbation theory result.
For the $\sigma(\mu^+ \mu^-)$ here the corresponding reduction
from perturbation theory is even greater, 22\% at RHIC and 14\% LHC.
The present perturbative $\sigma(\mu^+ \mu^-)$ calculations are in fairly
good agreement with
the calculations of Hencken, Kuraev, and Serbo\cite{hks},
but the present exact cross section calculations are in disagreement
with their argument that
Coulomb corrections are relatively insignificant for $\mu$ pairs.

So far the calculations presented have been performed in the impact parameter
independent representation.  There is an additional reduction that comes into
play for an observable $\sigma(\mu^+ \mu^-)$ that arises from unitarity
considerations, and one must make use of the impact parameter
representation discussed in the following section.

\section{Impact parameter and unitarity}

The perturbative (Born) cross section and corresponding cross sections with
higher order Coulomb corrections discussed in the previous section
correspond to an inclusive cross section, constructed from a probability
corresponding to the number operator for a given process.  If one considers
an exclusive cross section, e.g. exciting a $\mu$ pair and
nothing else in a heavy ion reaction, then one must consider unitarity
corrections for competing processes in an impact parameter representation
as will be seen below. For
$\mu$ pair production the main unitarity corrections arise in principle
from competing $e^+ e^-$ pair production, Coulomb dissociation of the
heavy ions, and nuclear processes at ion-ion overlap.

\begin{figure}[h]
\begin{center}
\epsfig{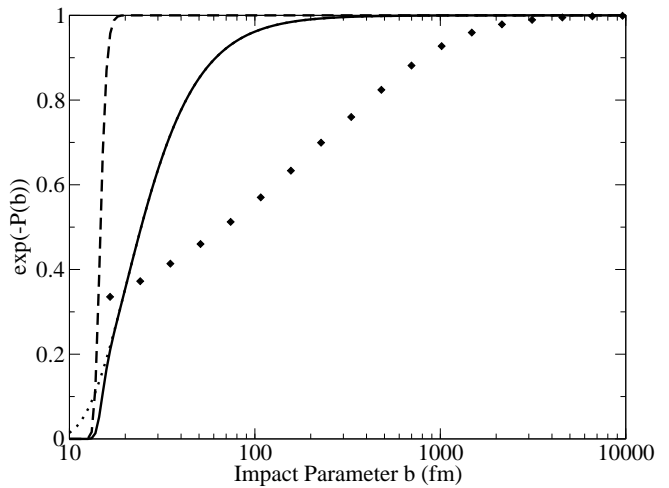}
\end{center}
\caption{Unitarity reduction factor for relevant processes in the calculation
of exclusive $\mu$ pair production at RHIC: dashed line -- nuclear collisions;
dotted line -- Coulomb excitation of one of the heavy ions; solid line --
nuclear collision or Coulomb excitation;
diamonds -- $e^+ e^-$ pair production.}
\end{figure}

Hencken, Kuraev and Serbo have observed that while unitarity corrections
are small for $e^+ e^-$ cross sections they are large for corresponding
$\mu^+ \mu^-$ pair production\cite{hks}. The perturbative Born cross section
for $e^+ e^-$ production, corresponding to an inclusive cross section, is
little increased from the exclusive cross section.
The perturbative Born cross section for $\mu^+ \mu^-$ production also
corresponds to the inclusive cross section, but the exclusive cross
section is significantly reduced by unitarity corrections
due to the simultaneous production of $e^+ e^-$ pairs along with the 
$\mu^+ \mu^-$ pairs.  Lee and Milstein have recently developed a
quasi-analytical procedure to include the higher order Coulomb
corrections in calculating the impact parameter dependence the $e^+ e^-$
pair production\cite{lm3}.
Based on their procedure Jenschura, Hencken and Serbo have updated
the consideration of the $e^+ e^-$ pair unitarity corrections\cite{jhs}.

In practice some unitarity corrections are relevant to what is actually
measured and some are not.  While it is an enlightening theoretical exercise
to consider $e^+ e^-$ pair unitarity corrections to $\mu$ pair rates, in
practice the dominant contributions of soft $e^+ e^-$ pairs are of an energy
scale orders of magnitude too small to be observed in an experiment designed
to observe $\mu$ pairs.  On the other hand, when calculating $\mu$ pair rates
in an impact parameter representation, a correction must be made to exclude
the lowest impact parameters of ion-ion overlap, where the dominant processes 
are nuclear.

\begin{figure}[h]
\begin{center}
\epsfig{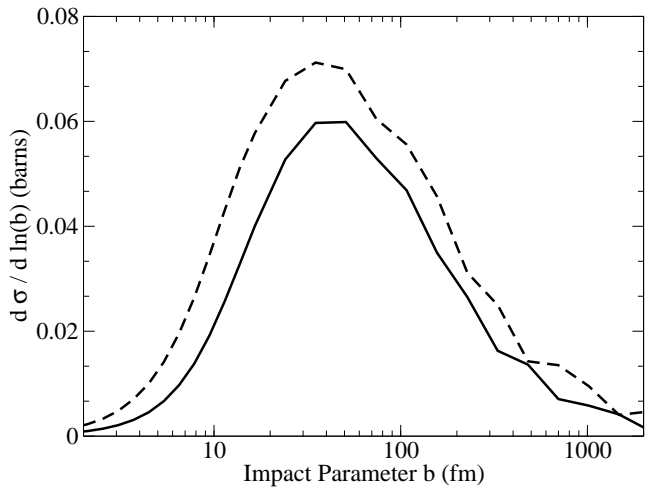}
\end{center}
\caption{Impact parameter dependence of contribution to total cross section
for $\mu$ pair production at RHIC: dashed line -- perturbation theory;
solid line -- higher order calculation.}
\end{figure}

If one assumes independence of the various heavy ion reaction processes, then
the probability of a reaction leading to a final state differing from the
incoming channel is given by the usual Poisson distribution.
If $P(b)$ is the sum of the probabilities for producing an excited state by
a heavy ion reaction at a given impact parameter $b$ 
\begin{equation}
P(b)=\Sigma_i P_i(b),
\end{equation}
where each $P_i(b)$ is the inclusive probability of a specific final state $i$,
then the exclusive probability for a given final state $P_i^u(b)$ is
\begin{equation}
P_i^u(b)=P_i(b) exp(-P(b)),
\end{equation}
and the probability of remaining in the initial state is
\begin{equation}
P_o^u(b)=1 -  exp(-P(b)).
\end{equation}

Figure 1 shows the unitarity reduction factor $exp(-P(b))$ in Eq. (12)
evaluated for the probabilities of various processes for the case of
Au + Au at RHIC.  In agreement with the previously discussed above work in the
literature\cite{hks,lm3,jhs} the unitarity
effect of $e^+ e^-$ pair production (diamonds) is significant for low
and intermediate impact parameters.  Following the methods of
Ref.\cite{brw,bcw} I have also calculated the unitarity reduction factor
$exp(-P(b))$ for Coulomb dissociation  and nuclear dissociation.

It is instructive to compare the impact parameter dependence of the
contribution to $\mu$ pair production and $e^+ e^-$ pair production
at RHIC.  Figure 2 shows the distribution for for $\mu$ pairs and Figure 3
for $e^+ e^-$ pairs.  In both figures dashes correspond to perturbation
theory and the solid line the higher order calculation. The shift in scale
mentioned in the previous section is evident.  
Comparing the region of $e^+ e^-$ reduction shown in Fig.1 with the regions
of dominant cross section contribution for $\mu$ pairs (Fig. 2) and $e^+ e^-$
pairs (Fig. 3) makes evident the reasoning of Hencken, Kuraev and
Serbo\cite{hks} that unitarity corrections
are small for $e^+ e^-$ cross sections and large for
$\mu^+ \mu^-$. Also clearly the region
of nuclear collisions (dashed line in Fig. 1) would provide no reduction of the
$e^+ e^-$ pair cross section (Fig. 2), but would slightly reduce the
$\mu$ pair cross section (Fig. 1). It is obvious that even with a momentum
dependent form factor there is significant contribution to the $\mu$ pair
cross section here in the region of ion-ion overlap.
This contribution must be eliminated for ultraperipheral collisions, and
leads into the discussion in the rest of this section.

\begin{figure}[h]
\begin{center}
\epsfig{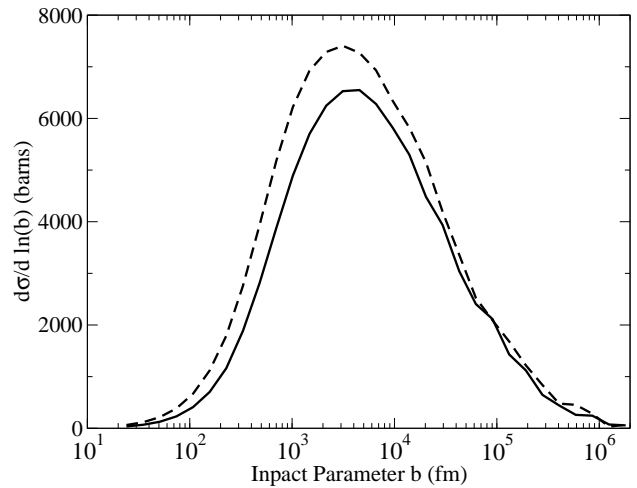}
\end{center}
\caption{Impact parameter dependence of contribution to total cross section
for $e^+ e^-$ pair production at RHIC: dashed line -- perturbation theory;
solid line -- higher order calculation.}
\end{figure}

As noted above, what states are considered in the unitarity consideration
can be determined or defined by the energy scale of the detected particles.
For example, when considering $\mu^+ \mu^-$ pair production, one
might not consider the dominant soft $e^+ e^-$ pairs as part of the
the excited spectrum for purposes of unitarity normalization.  This might
be a reasonable definition corresponding to the experimental detection
conditions. On the other hand,
to construct a calculated cross section corresponding to the observed
pair production events without any other final state particles arising
from ion-ion overlap, one should include only ultraperipheral
impact parameters.

In terms of a $\bf{b}$ (impact parameter) dependent
amplitude $M(\bf{k,b})\exp(\it{i} \bf{k \cdot b})$ an appropriate
non-unitarized probability can be written
\begin{equation}
P_i(b) = \vert \int d^2 k M_i(\bf{k,b}) \exp(\it{i} \bf{k \cdot b})\vert^2.
\end{equation}
Let $P_i(b)$ be a non-unitarized probability for exciting a $\mu^+ \mu^-$
pair and $P_j(b)$ the corresponding probability for a nuclear reaction.
Then define a partially exclusive cross section as one that excludes nuclear
interaction only
\begin{equation}
\sigma_i = \int_0^\infty d^2  b P_i^u(b) = 
\int_0^\infty d^2  b P_i(b) exp(-P_j(b)).
\end{equation}
Since the nuclear interactions occur only below some $b_0 = R_1 + R_2$,
 is convenient to express this cross section as a difference
\begin{eqnarray}
\sigma_i&=&\int_0^\infty d^2  b P_i(b)
\nonumber \\
&+&\int_0^{b < b_0} d^2  b  P_i(b) (exp(P_j(b) - 1 )).
\end{eqnarray}
The first term can be evaluated as was done before without a specific impact
parameter representation\cite {ajb05}.  The second term can then be evaluated
using the method of Ref.\cite {ajb06}: since $\bf{b}$ is limited to the
lowest impact parameters the $\exp(\it{i} \bf{k \cdot b})$ factor is still
numerically tractable even though $\bf{k}$ scales up by a factor
of $m_{\mu}/m_e$ as compared to the $e^+ e^-$ case.  A similar trick to
subtract the nuclear interaction at small impact parameters has previously
been used\cite{aktb}.

\section{Numerical results}

Table I summarizes the total cross section results for $\mu^+ \mu^-$ pair
production at RHIC.  In the first row the $\bf{b}$-independent perturbative
cross section for $\sigma(\mu^+ \mu^-)$ is 211 mb. and the cross section with
higher order effects is 164 mb.  As previously noted in Section III, this
reduction of 22\% from perturbation theory is greater than the 17\% reduction
in the exact $\sigma(e^+ e^-)$ from the perturbation theory seen in
Ref.\cite{ajb05}.  The second row shows the results from integrating
the $\bf{b}$-dependent computation of the cross section shown in Fig. 2.
The numbers in parentheses correspond to the negative of the second
right hand term of Eq. (16). 
A rough check can be done by comparing the results of
the $\bf{b}$-dependent and $\bf{b}$-independent calculations of the total
$\mu^+ \mu^-$ production cross section.  Even though the $\bf{b}$-dependent
becomes more inaccurate beyond the low impact parameters it still reproduces
the $\bf{b}$-independent results to about 10\%.  The lowest impact parameters
have the greatest accuracy and we calculate the higher order cross section
from the overlap impact parameters to be subtracted off as 20 mb.  Thus
the final computed best cross section (third row) is 144 mb, a 32\% reduction
from the perturbation theory calculation.
\begin{table}
\caption[Table I]{RHIC: Au + Au, $\gamma = 100, \mu^+ \mu^-$ total
cross section.}
\begin{tabular}{|c|cc|}
\colrule
 RESULTS IN mb & Perturb. & Exact  \\
\colrule
b independent formulation  & 211 & 164 \\
b integration (ion overlap) & 232 (36) & 181 (20) \\
b independent minus ion overlap & 175 & 144 \\
Hencken et al. ($\gamma = 108$) & 230 & 230 \\
\colrule
\end{tabular}
\label{tabi}
\end{table}
\begin{table}
\caption[Table II]{LHC: Pb + Pb, $\gamma = 2760, \mu^+\mu^-$ total cross section.}
\begin{tabular}{|c|cc|}
\colrule
 RESULTS IN barns & Perturb. & Exact  \\
\colrule
b independent formulation  & 2.43 & 2.09 \\
b integration ion overlap & 0.13 & 0.06 \\
b independent minus ion overlap & 2.29 & 2.03 \\
Hencken et al. ($\gamma = 3000$) & 2.60 & 2.60 \\
\colrule
\end{tabular}
\label{tabj}
\end{table}

Table II shows calculations of colliding Pb + Pb ions at the LHC.
The perturbative $\mu^+ \mu^-$ production cross section shown in the
first row is 2.43 b and higher order effects reduced it by 14\% to 2.09 b.
Again this reduction is greater than the 11\%  reduction in the
exact $\sigma(e^+ e^-)$ from the perturbation theory result\cite {ajb05}.
Due to the higher values of transverse momentum transferred from the virtual
photons in this LHC case it was not feasible to compute the $\bf{b}$-dependent
cross section contributions throughout the entire impact parameter range.
However in the region of ion overlap the impact parameter was small enough
that the the a rapidly oscillating phase
$\exp(i \bf{k \cdot b})$ in the integral over the transverse momentum $\bf{k}$
transfered from the ion to the lepton pair remained tractable and
Eq. (16) could be utilized.
The additional reduction from exclusion of overlap impact parameters
was 0.06 b for a best value of 2.03 b, an overall 16\% reduction from
perturbation theory.

The present perturbative $\sigma(\mu^+ \mu^-)$ calculations are in fair
agreement with the calculations of Hencken, Kuraev, and Serbo\cite{hks},
but the present exact cross section calculations are in disagreement
with their argument that
Coulomb corrections are relatively insignificant for $\mu$ pairs.  In this work
I have shown that unlike the case for $e^+ e^-$ pair production, the finite
size of the colliding nuclei provides an important modification for both
the perturbative and higher order calculated total cross sections
$\sigma(\mu^+ \mu^-)$.  The form factor reduces the higher order calculation
by an even greater percentage than it does for perturbation theory.
Furthermore, making the necessary elimination of interactions where the ions
overlap further reduces the higher order cross section from perturbation
theory.

\section{Acknowledgment}

This manuscript has been authored
under Contract No. DE-AC02-98CH10886 with the U. S. Department of Energy.

\appendix
\section{Higher order form factor effects}
As noted in Section III, to include a form factor in the
eikonalized expression for the transverse integral with
Coulomb corrections, the most obvious prescription is to apply
the form factor to the transverse potential, leading to Eq. (9):
\begin{equation}
F^f({\bf k}) = 2 \pi \int d \rho \rho J_0(k \rho)
\{\exp [2i Z\alpha g(k) K_0(\rho \omega / \gamma)] -1 \} .
\end{equation}
But including the form factor in the transverse potential is equivalent to
letting
the coupling constant $Z \alpha$ run as a function of $k$, analogous to the
situation in QCD.  To do this makes the numerical integration of Eq. (1)
more complicated and has not
been done in this paper.  However, by a relatively simple
modification of the $b$ independent expression for the higher order cross
section one can put an upper limit on the modification to higher order
effect of using the more proper Eq. (9) rather than the expression Eq. (10)
utilized throughout this this paper.

The numerical integration of Eq. (1) is most conveniently carried out
after a change of variables to $\xi = k \rho$ and
\begin{equation}
F({\bf k}) =  {2 \pi \over k^2}
\int d \xi \xi J_0(\xi) \{\exp [2i Z
\alpha K_0(\xi \omega / \gamma k)] - 1 \}.
\end{equation}
This integral is carried out for various values of the parameter
$k \gamma / \omega$. $F$ actually has a two dimensional parameterization
in $k \gamma / \omega$ and $k$, but the $1/k^2$ dependence trivially factors
out of the integral.  Likewise the additional $k$ dependence
in expression Eq. (10) utilized in this paper factors out trivially.
However the additional $k$ dependence in the more exact expression Eq. (9)
does not factor out trivially, leading to the additional complication. 

To put a limit on the error incurred by using the expression Eq. (9) rather
than Eq. (10), I begin by recalling that the organization of the
$b$-independent computer code utilized in Ref. \cite{ajb05} involves a
difference
\begin{equation}
\vert \Delta F({\bf k})^2 \vert = \vert F({\bf k}) \vert^2
- \vert F_0({\bf k}) \vert^2
\end{equation}
between the squared value of the higher order expression Eq. (1) and that of
the perturbative expression, Eq. (2).  Since the effect
of the form factor $g(k)$ is the same as reducing the value of $Z$ as a
function of $k$ at large $k$ (like running coupling in QCD) then evaluation
of Eq. (A2) for various values of $Z$ may be used as a proxy for the
higher order dependence on $g(k)$.  That is, for a given $Z$ if the form
factor is reduced from unity by some percentage then it is equivalent to
no form factor and just reducing $Z$ by the same percentage.
Numerical calculations of $\vert \Delta F^({\bf k})^2 \vert$ show that it
scales as $Z^4$ for low values of $Z$ and a little less than $Z^4$ as $Z$ is
increased.  This scaling is consistent with the integral over (A3)
\begin{equation}
G = \int { d^2 k \over ( 2 \pi )^2 } k^2 [ \vert F({\bf k}) \vert^2
- \vert F_0({\bf k}) \vert^2 ]
\end{equation}
in Lee and Milstein's analysis of higher order Coulomb
corrections\cite{lm1,lm2}, which takes the analytical form
\begin{equation}
G = -8 \pi (Z \alpha)^2 [Re \psi ((1 + i Z \alpha) + \gamma_{Euler}],
\end{equation}
where $\psi ((1 + i Z \alpha)$ is the digamma function and $\gamma_{Euler}$ is
Euler's constant.  This expression may be alternatively expressed as
\begin{equation}
G = -8 \pi (Z \alpha)^2 f(Z \alpha),
\end{equation}
where $f(Z \alpha)$ is the same function that was presented by Bethe,
Maximon and Davies\cite{bem}
for Coulomb corrections to $e^+ e^-$ photoproduction on heavy nuclei
and takes the form
\begin{equation}
f(Z \alpha)= (Z \alpha)^2 \sum_{n=1}^\infty {1 \over n(n^2+(Z \alpha)^2)}.
\end{equation}
$G$ obviously scales as $Z^4$ for low values of $Z$ and a little less than
$Z^4$ as $Z$ is increased.

The expression Eq. (A3) modified with the lowest order implementation of
the form factor utilized in this paper takes the form
\begin{eqnarray}
\vert \Delta F^{f0}({\bf k})^2 \vert &=& \vert F^{f0}({\bf k}) \vert^2
- \vert F_0^{f0}({\bf k}) \vert^2
\nonumber \\
&=& g(k)^2 ( \vert F({\bf k}) \vert^2
- \vert F_0({\bf k}) \vert^2 ).
\end{eqnarray}
If one consider the analogous expression with the form factor to higher order,
then replacing the $g(k)^2$ dependence of Eq. (A7) with $g(k)^4$ suggested by
the $Z^4$ scaling seen in the difference without a form factor,
\begin{eqnarray}
\vert \Delta F^{f}({\bf k})^2 \vert &=& \vert F^{f}({\bf k}) \vert^2
- \vert F_0^{f}({\bf k}) \vert^2
\nonumber \\
&=& g(k)^4 ( \vert F({\bf k}) \vert^2
- \vert F_0({\bf k}) \vert^2 ),
\end{eqnarray}
should slightly overstate the higher order effect of the form factor in Coulomb
corrections.

Recalculation of the exact $b$ independent RHIC Au + Au cross section of
Table I
with the $g(k)^4$ form factor scaling of Eq. (A9) gives 171 mb, a 19\%
reduction from perturbation theory in comparison with the 164 mb 22\% reduction
using the more approximate $g(k)^2$ of Eq. (A8).  Likewise for Pb + Pb at
LHC the exact calculation with $g(k)^4$ gives
 2.12 barns, a 13\% reduction from
perturbation theory in comparison with the 2.09 barn 14\% reduction with
$g(k)^2$.

Both recalculations make only a small change from the lowest order treatment
of the form factor in this paper.  And since it is far from trivial to
implement the more proper higher order treatment of the form factor of Eq. (9),
especially in $b$ dependent calculations, I have not done so in this paper.
It seems that once the $k$ dependent cutoff of the form factor is put in,
then sharpening the cutoff by an additional squaring has a relatively small
effect.  Even with the higher order $g(k)^4$ scaling calculations, the
reduction from perturbation theory in $\sigma(\mu^+ \mu^-)$ are still larger
than the reductions
in the analogous exact $\sigma(e^+ e^-)$ without a form factor from
perturbation theory.

\section{Longitudinal form factor effects}
One might include longitudinal form factor effects by modifying Eq. (3)
to make $g(k)$ a function of $k^2 + \omega^2 / \gamma^2$ rather than
simply a function of $k^2$:
\begin{equation}
g(k) = {1 \over 1 + (k^2 + \omega^2/ \gamma^2) / \Lambda^2}.
\end{equation}
I have recalculated $b$ independent cross sections using Eqns. (5),(10)
and (B1) in place of (3),
and  I find a 5\% reduction for both the perturbative and higher order
computations for RHIC but only a corresponding 1\% reduction for LHC.  The
5\% reduction for RHIC is equivalent to calculations without
a longitudinal factor, but with the value of $\Lambda$ reduced from 80 MeV
to 75.5 MeV.

\end{document}